\documentclass[nonacm]{acmart}
\usepackage{chronology}
\AtBeginDocument{%
  \providecommand\BibTeX{{%
    \normalfont B\kern-0.5em{\scshape i\kern-0.25em b}\kern-0.8em\TeX}}}

\setcopyright{none}
\copyrightyear{}
\acmYear{}
\acmDOI{}

\acmJournal{CSUR}
\acmVolume{}
\acmNumber{}
\acmArticle{}
\acmMonth{}

\begin{document}
\title{A Survey on Locality Sensitive Hashing Algorithms and their Applications}
\author{Omid Jafari}
\email{ojafari@nmsu.edu}
\orcid{0000-0003-3422-2755}
\affiliation{%
  \institution{New Mexico State University}
  \department{Computer Science}
  \city{Las Cruces}
  \state{NM}
  \country{USA}
}

\author{Preeti Maurya}
\email{preema@nmsu.edu}
\orcid{0000-0000-0000-0000}
\affiliation{%
	\institution{New Mexico State University}
	\department{Computer Science}
	\city{Las Cruces}
	\state{NM}
	\country{USA}
}

\author{Parth Nagarkar}
\email{nagarkar@nmsu.edu}
\orcid{0000-0001-6284-9251}
\affiliation{%
	\institution{New Mexico State University}
	\department{Computer Science}
	\city{Las Cruces}
	\state{NM}
	\country{USA}
}

\author{Khandker Mushfiqul Islam}
\email{mushfiq@nmsu.edu}
\orcid{0000-0000-0000-0000}
\affiliation{%
	\institution{New Mexico State University}
	\department{Computer Science}
	\city{Las Cruces}
	\state{NM}
	\country{USA}
}

\author{Chidambaram Crushev}
\email{crushev@nmsu.edu}
\orcid{0000-0000-0000-0000}
\affiliation{%
	\institution{New Mexico State University}
	\department{Computer Science}
	\city{Las Cruces}
	\state{NM}
	\country{USA}
}

\renewcommand{\shortauthors}{Jafari, et al.}
\begin{abstract}
Finding nearest neighbors in high-dimensional spaces is a fundamental operation in many diverse application domains. Locality Sensitive Hashing (LSH) is one of the most popular techniques for finding approximate nearest neighbor searches in high-dimensional spaces. The main benefits of LSH are its sub-linear query performance and theoretical guarantees on the query accuracy. In this survey paper, we provide a review of state-of-the-art LSH and Distributed LSH techniques. Most importantly, unlike any other prior survey, we present how Locality Sensitive Hashing is utilized in different application domains.
\end{abstract}

\begin{CCSXML}
	<ccs2012>
	<concept>
	<concept_id>10002944.10011122.10002945</concept_id>
	<concept_desc>General and reference~Surveys and overviews</concept_desc>
	<concept_significance>500</concept_significance>
	</concept>
	</ccs2012>
\end{CCSXML}

\ccsdesc[500]{General and reference~Surveys and overviews}

\keywords{Locality Sensitive Hashing, Approximate Nearest Neighbor Search, High-Dimensional Similarity Search, Indexing}

\maketitle

\section{Introduction}
\label{sec:introduction}
Finding nearest neighbors in high-dimensional spaces is an important problem in several diverse applications, such as multimedia retrieval, machine learning, biological and geological sciences, etc. For low-dimensions ($<10$), popular tree-based index structures, such as KD-tree \cite{bentley1975multidimensional}, SR-tree \cite{katayama1997sr}, etc. are effective, but for higher number of dimensions, these index structures suffer from the well-known problem, \textit{curse of dimensionality} (where the performance of these index structures is often out-performed even by linear scans) \cite{chavez2001searching}. Instead of searching for exact results, one solution to address the \textit{curse of dimensionality} problem is to look for \textit{approximate} results. In many applications where 100\% accuracy is not needed, searching for results that are \textit{close enough} is much faster than searching for exact results \cite{datar2004locality}. Approximate solutions trade-off accuracy for a much faster performance. The goal of the approximate version of the nearest neighbor problem, also called \textit{c-approximate Nearest Neighbor search}, is to return objects that are within $c \times R$ distance from the query object (where $c>1$ is a user-defined approximation ratio and $R$ is the distance of the query object from its nearest neighbor). 

\subsection{Locality Sensitive Hashing}
Locality Sensitive Hashing \cite{datar2004locality} is one of the most popular solutions for the approximate nearest neighbor (ANN) problem in high-dimensional spaces. Locality Sensitive Hashing (LSH) maps high-dimensional data to lower dimensional representations by using \textit{random} hash functions. Data points are assigned to individual hash buckets in each hash function. The idea behind this approach is that closer data points in the original high-dimensional space will be mapped to the same hash buckets in the lower-dimensional projected space with a high probability. Since it was first introduced in \cite{gionis1999similarity}, many variants of Locality Sensitive Hashing have been proposed \cite{bawa2005lsh, lv2006time, zhang2010data, gan2012locality, liu2014sk, sun2014srs, huang2015query, liu2019lsh} that mainly focused on improving the search accuracy and/or the search performance of the given queries. The performance/accuracy trade-off of the query is determined by a user-provided \textit{success guarantee} (where a high success guarantee will return a result with high accuracy at the expense of faster performance and vice-versa). 

\subsection{Motivation}
\label{sec:motivation}
\subsubsection{Motivation for using LSH:}\hspace*{\fill} \\
Locality Sensitive Hashing (LSH) is known for two main advantages: its sub-linear query performance (in terms of the data size)
and theoretical guarantees on the query accuracy. Additionally, LSH uses random hash functions which are data-independent (i.e. data properties such as data distribution are not needed to generate these random hash functions). Additionally, the data distribution does not affect the generation of these hash functions. Hence, in applications where data is changing or where newer data is coming in, these hash functions do not require any change during runtime. 
While the original LSH index structure suffered from large index sizes (in order to obtain a high query accuracy) \cite{bawa2005lsh, lv2006time}, state-of-the-art LSH techniques \cite{gan2012locality, huang2015query} have alleviated this issue by using advanced methods such as \textit{Collision Counting} and \textit{Virtual Rehashing}. Hence, owing to their small index sizes, fast index maintenance, fast query performance, and theoretical guarantees on the query accuracy, Locality Sensitive Hashing is still considered an important technique for solving the Approximate Nearest Neighbor problem. 

\subsubsection{Motivation of our work (difference with other surveys):}\hspace*{\fill} \\
LSH-based algorithms are used in several application domains such as content-based multimedia retrieval systems, computational biological/medical studies, earth sciences, information retrieval tasks, etc. Most of these works are basing their methods on the original Euclidean distance based LSH (E2LSH \cite{datar2004locality}). However, E2LSH has several drawbacks and one goal of this survey paper is to show the workflow of several other LSH-based algorithms. This work will help the application domains to improve their efficiency by changing their base algorithm. 

There has been several surveys on approximate nearest neighbor search methods, such as \cite{wang2014hashing, chi2017hashing, wang2017survey, aumuller2017ann, li2019approximate, cai2019revisit}, that have reviewed some of the hashing-based algorithms. In \cite{wang2014hashing}, authors review hashing-based and quantization-based methods in solving similarity search problems. Moreover, for each of the methods, various aspects such as the hash functions, distance measures, and search techniques are also reviewed. \cite{chi2017hashing} reviews hashing-based techniques used in domains such as information systems. Moreover, it categorizes the techniques in two major groups: data-oriented hashing and security-oriented hashing. \cite{wang2017survey} reviews the hashing-based methods and specifically the learning to hash and quantization-based solutions to solve similarity search problems. Learning to hash methods are data-dependent techniques that aim to learn hash functions from a specific given dataset. \cite{aumuller2017ann} presents a tool for benchmarking in-memory approximate nearest neighbor search algorithms. Moreover, several graph-based, tree-based, and LSH-based algorithms are experimentally compared using real datasets. In \cite{li2019approximate}, authors conduct an experimental study over several LSH-based, learning to hash, partition-based, and graph-based algorithms. Another experimental study is presented in \cite{cai2019revisit} that compares tree-based, hashing-based, quantization-based, and graph-based methods using real datasets.

As mentioned earlier, the previous works have reviewed some of the LSH-based techniques; however, they do not provide an extensive review on LSH-based techniques. Different from the previous works, in this survey paper, we focus only on LSH-based techniques to solve the ANN problem, and we review the latest advances in LSH-based techniques. Moreover, in this survey paper, we review two aspects that, to the best of our knowledge, are not included in any other survey papers; Distributed LSH frameworks and applications of LSH-based techniques in various diverse domains.

\subsection{Contributions}
\label{sec:contributions}
In this paper, we present an in-depth review of the recent advances in Locality Sensitive Hashing techniques. Our contributions are listed as following:
\begin{itemize}
	\item We perform an in-depth review over LSH-based techniques by categorizing them based on the hash family that they use and explaining their work flow.
	
	\item There are different distributed frameworks proposed to improve the processing speed of the LSH algorithms. In this paper, we review these distributed frameworks and present an overview of their architecture.

	\item LSH-based algorithms are used in various application domains. In this survey, we categorize the application domains and explain how LSH-based algorithms are utilized in each of them.
\end{itemize}

\subsection{Paper Organization}
\label{sec:organization}
The remainder of the paper is organized as follows: In section \ref{sec:background}, we provide background information related to LSH. Section \ref{sec:lshTechniques} presents a detailed review of LSH-based algorithms that are proposed to solve and improve the approximate nearest neighbor search problem. Section \ref{sec:distributed} presents a detailed review of the distributed frameworks for LSH-based algorithms. In Section \ref{sec:applications}, we present the works in different application domains that use Locality Sensitive Hashing. Finally, we conclude the paper in Section \ref{sec:conclusions}.

\section{Background and Key Concepts}
\label{sec:background}
In this section, we describe the key concepts behind Approximate Nearest Neighbor (ANN) search. Note that, while there are several space partitioning and graph-based methods that also tackle the ANN problem, our focus in this paper is specifically on Locality Sensitive Hashing-based methods. We refer the reader to \cite{chi2017hashing} for discussions on non-LSH based methods. 

\noindent Given a dataset $\mathcal{D}$ with $n$ points and $d$ dimensions and a query point $q$ in the same space as the dataset, the goal of $c$-ANN search (where $c>1$ is an approximation ratio) is to return points $o \in \mathcal{D}$ such that $dist(o, q) \leq c \times dist(o^*, q)$, where $o^*$ is the true nearest neighbor of $q$ in $\mathcal{D}$ and $dist$ is the distance between the two points. Similarily, $c$-$k$-ANN search aims at returning top-$k$ points such that $dist(o_i, q) \leq c \times dist(o_i^*, q)$, where $1 \leq i \leq k$. 

\noindent Hashing-based methods try to find the nearest neighbors in high-dimensional datasets by projecting them into one or more low-dimensional spaces using hash functions. LSH is a famous hashing-based method that creates the low-dimensional projections such that the localities of the original space are preserved in them (i.e. two nearby points in the original space are also nearby in the projected space). 

\noindent For two points $x$ and $y$ in a $d$-dimensional dataset $D \subset \mathbb{R}^d$, we say a hash function $H$ is ($R$, $cR$, $p_1$, $p_2$)-sensitive if it satisfies the following two conditions:

\begin{itemize}
	\item if $|x - y| \leq R$, then $Pr[H(x) = H(y)] \geq p_1$, and
	\item if $|x - y| > cR$, then $Pr[H(x) = H(y)] \leq p_2$
\end{itemize}

Here, $c$ is an approximation ratio and $p_1$ and $p_2$ are probabilities. In order for this definition to work, $c  > 1$ and $p_1 > p_2$. The definition states that two points $x$ and $y$ are hashed to the same bucket in the projection with a very high probability $\geq p_1$ if they are close to each other, and if they are not close to each other, then they will be hashed to the same bucket with a low probability $\leq p_2$. Next, we present the popular hash function families for the Hamming, Minkowski, Angular, and Jaccard distances. 

\noindent For the Hamming metric, \cite{indyk1998approximate} defined the LSH function as $H(x) = x_i$, where $x_i$ is the $i$-th dimension of the point $x$ ($i \in [1, d]$). Therefore, for two points $x$ and $y$ with a Hamming distance of $r$, the probability that they have the same hash value is $Pr[H(x) = H(y)] = 1 - \frac{r}{d}$.

\noindent For the Minkowski metric, \cite{datar2004locality} defined the LSH functions as $H_{\vec{a},b} (x) = \left\lfloor{\frac{\vec{a}.x + b}{w}}\right\rfloor$, where $\vec{a}$ is a $d$-dimensional random vector chosen from the standard $p$-stable distribution and $b$ is a real number chosen uniformly from $[0, w)$, such that $w$ is the width of the hash bucket. To generate $\vec{a}$, Cauchy, Gaussian (Normal), and L\'evy distributions are used for $p=1$, $p=2$, and $p=\frac{1}{2}$ respectively. Therefore, for two points $x$ and $y$ with a Minkowski distance of $r$, the probability that they have the same hash value is $Pr[H(x) = H(y)] = \int_{0}^{w}{f_p(\frac{t}{r})(1-\frac{t}{w}) dt}$. Here, $f_p(\frac{t}{r})$ is the density function of the $p$-stable distribution.

\noindent For the Angular metric, \cite{charikar2002similarity} defined the LSH functions as $H(x) = sgn(\vec{a} . x)$, where $sgn$ is the sign function and $\vec{a}$ is a random vector drawn from the Normal distribution. In this case, for two points $x$ and $y$ with $\theta$ defined as the angle between them, the probability that they have the same hash value is $Pr[H(x) = H(y)] = 1 - \frac{\theta}{\pi}$.

\noindent Finally, for the Jaccard metric, \cite{broder2000min} defined the LSH functions as $H(x) = min \{ \pi(x_i) \}$, where $x_i \in x$ and $\pi$ is a random permutation from the set of all possible permutations. Therefore, for two points $x$ and $y$ with a Jaccard similarity of $J$, the probability that they have the same hash value is $Pr[H(x) = H(y)] = J$.

\section{Review of LSH Techniques}
\label{sec:lshTechniques}
In this section, we present our detailed review of LSH techniques. The main benefits of LSH are its sub-linear query time and the theoretical guarantees provided on the query accuracy. We classify different LSH techniques based on the distance hash function family (Section \ref{sec:background}), in particular, Hamming distance, Minkowski distance, Angular distance, and Jaccard distance. 

\subsection{Hamming-based LSH Techniques} % SUBSECTION

Locality Sensitive Hashing was first proposed in \cite{indyk1998approximate} for the Hamming distance to solve the $(R, c)$-near neighbor search problem. The proposed method uses multiple hash functions and hash tables to be able to guarantee a good search quality. Moreover, authors theoretically find the optimal number of hash functions and hash tables in order to have constant hashing probabilities.

\noindent Boosted LSH (BLSH) is proposed in \cite{kim2020boosted}. This method trains linear classifiers sequentially using an adaptive boosting paradigm that results in reducing the redundancy in LSH projections. Further, BLSH is experimentally shown to have comparable performance against other machine/deep learning approaches in the speech enhancement applications.

\subsection{Minkowski-based LSH Techniques} % SUBSECTION

E2LSH \cite{datar2004locality} is the first LSH method that is proposed for the Euclidean distance. The main idea of E2LSH is to use multiple hash tables and compound hash functions to increase the collision chance of two nearby points. By using multiple hash tables and multiple hash functions, E2LSH reduces the number of false positives and false negatives while keeping the accuracy high. In \cite{lv2006time}, hash-perturbation LSH is proposed that perturbs the hash values on the projections to reduce the large number of required hash tables in the basic LSH. Multi-probe LSH \cite{lv2007multi} uses the intuition that nearest neighbors are more likely to be hashed to the close-by buckets and intelligently looks into multiple neighboring buckets. Moreover, it assigns a distance score to each bucket, and later, the buckets are accessed in the increasing order of their scores. By using this strategy, Multi-probe requires less number of hash tables while achieving the same accuracy. Authors in \cite{joly2008posteriori} improve on the multi-probe LSH \cite{lv2007multi} and utilize probabilistic approaches instead of likelihood. This is done via estimating the distribution of the neighboring points of a query. Using this approach, a probabilistic score is created that is used in the multi-probe search. In \cite{jegou2008query}, authors propose a query adaptive hashing method that in the query processing phase, it uses the expected accuracy measurement to choose the best hash functions that are more appropriate for the given query. By using this query adaptive strategy, the proposed technique improves the accuracy of the results.

\noindent In \cite{cayton2008learning}, the authors propose a framework that learns from data-set characteristics (i.e. density) to intelligently choose LSH hash functions and thus, improve the accuracy. The idea is to have fine/smaller buckets in the dense areas of the data-set and coarse/larger buckets in the sparse areas. To this end, dynamic programming is used at each data-set dimension to repeatedly divide the data-set points into left and right bins until the desired size is reached. \cite{zhang2010data} proposes a data dependent technique that uses Principal Components Analysis to lower the dimensions of the dataset such that a uniform dataset is generated. The proposed method generates projections that are more uniform than the random projections. Authors in \cite{dasgupta2011fast} use Hadamard transformations to better estimate the distances in the Euclidean and angular spaces and reduce the running time of LSH methods. They propose two methods called ACHash and DHHash, where the former uses one Hadamard transformation and the latter uses two Hadamard transformations. In C2LSH \cite{gan2012locality}, there is only one hash table and $m$ random hash functions (also called projections). Each projection is divided into buckets of width $w$ and data-set points are hashed into each projection using the hash functions. Moreover, an approach called ``collision counting" is proposed that counts the number of times a hashed point is mapped to the same bucket as the query point and as soon as $l$ collisions are occurred for any point, that point is considered a candidate. C2LSH has two stopping conditions: 1) $k$ + $\beta n$ candidates are found, and 2) $k$ true positives are found. The true positives are found by calculating the Euclidean distance of the candidates with the query and checking if the distance is less than $cR$. If the stopping conditions are not met, the algorithm increases the projection search radius exponentially at each time (e.g. looking for collisions in two buckets instead of one bucket). 

\noindent Bi-level LSH \cite{pan2012bi} is proposed to improve accuracy and runtime of searches. It first partitions the dataset into random sub-groups using RP-tree; then, creates a single hash table for each sub-group and a hierarchical structure based on space-filling curves (Z-order curves). Boundary-expanding locality sensitive hashing \cite{wang2012boundary} works on the problem of nearest points being hashed into the boundaries of different buckets. It overcomes this problem by expanding the boundaries of each bucket; thus, increasing the collision probability of neighbors. The motivation of \cite{lee2013projection} is that sometimes points are projected to the boundaries of the buckets that will make them false negatives or false positives depending on the query bucket. To solve this issue, authors introduce the concept of projecting a point into more than one bucket using three hash functions that map the point to the 1) current bucket, 2) bucket to the left, and 3) bucket to the right. In \cite{gu2013improved}, the authors focus on the false positive removal process which requires Euclidean distance computation. To make the false positive removal faster in cases where there are many candidates generated, the authors use triangular inequality and a pivot-based algorithm to further prune the candidates. Finally, they experimentally show the speedup of their proposed method.

\noindent Dynamic Multi-probe LSH \cite{yin2013dynamic} is proposed to optimize I/O efficiency by dynamically changing the number of hash functions of each bucket such that the bucket fits a single disk page. A B+-tree is used for this process and as a result of this technique, buckets with varying granularity are generated. In \cite{zhang2013distribution}, a series of data-adaptive projections are generated using linear projections first to reduce the dimensionality of the dataset and then, using the distribution of the lower dimension data to generate the final hash functions. Further, an improved multi-probe strategy is proposed to improve the performance. In \cite{bai2014data}, a data-dependent approach is proposed for LSH. This approach, projects one dataset point to multiple random projections, and then, uses data distribution to learn a lower number of projections capable of approximating the projections created in the previous step. This approach is proposed for the Euclidean metric and experimentally shown to perform better than E2LSH. In \cite{xie2014data}, the K-means method is adopted to cluster the dataset into multiple groups and then E2LSH is performed on each cluster to construct LSH tables. Finally, it the benefits are experimentally shown.

\noindent In \cite{andoni2014beyond}, a data dependent hashing method is proposed. The proposed method uses two levels of hash functions to further prune the projections. Moreover, although the two-level hashing is data dependent, the hash functions are chosen randomly without any dependency to the data. Further, the time and space complexities of the proposed method are theoretically proven to be optimal. In \cite{wang2014bi}, authors use the same logic of Bi-level LSH and build a bi-level LSH method. During the first level, they use the K-means algorithm to partition the dataset into several clusters, and then, in the second level, they apply E2LSH \cite{datar2004locality} to each cluster. The motivation of this work is that items belonging to the same cluster will have a more uniform distribution; thus, E2LSH can perform better on them. SRS \cite{sun2014srs} uses R-tree index structures to estimate the original distances of the points from their projected distances. Moreover, it uses an incremental search strategy to look for nearest neighbors of a given query. The main contribution of SRS is to use small indexes while maintaining the same theoretical guarantees of  traditional LSH methods. SK-LSH \cite{liu2014sk} focuses on reducing the random I/Os by efficiently placing nearby projected points to the same or close disk pages. In order to do this, SK-LSH uses a space-filling curve and a new distance measure between the compound hash keys to estimate the distance between the points in the original space. A linear order is then used to sort the compound hash keys and store them on the disk.

\noindent QALSH \cite{huang2015query} creates query-aware hash functions with the intuition that when the query is near the bucket boundaries, its near neighbors might fall into another bucket, and in order to prevent that, QALSH considers the query point as the anchor of the buckets. Moreover, QALSH uses B+-trees for each hash function to improve the lookup time and make the range queries more efficient. LazyLSH \cite{zheng2016lazylsh} mentions that the chance of two nearby points being close in two different $l_p$ spaces is high. Therefore, in the indexing phase, LazyLSH builds an index for the $l_1$ space and call it a base space. In the query processing phase and when the query is in another $l_p$ space where $p \in (0, 1)$, it uses transformations to find nearest neighbors with a high accuracy. Authors in \cite{huang2017query}, extend QALSH \cite{huang2015query} to work with $l_p$ norms where $p \in (0, 2]$. They specifically use Levy distribution, Cauchy distribution, and Gaussian distribution for $l_{1/2}$, $l_1$, and $l_2$ norms respectively. Moreover, authors present a heuristic-based method called QALSH$^+$ that uses two-level indexing and KD-Tree structure to speed up the search process in datasets with large cardinalities.

\noindent I-LSH \cite{liu2019lsh} focuses on the radius expansion (Virtual Rehashing) process of LSH where the search radius in the projections are increased in order to look for candidates in the neighboring buckets of the query. Previous methods increase the radius exponentially; however, I-LSH uses an incremental way of increasing the radius. To incrementally increase the radius, I-LSH looks for the nearest point in the projection based on its distance to the query. This operation results in less disk I/Os since it prevents reading unnecessary buckets from the disk. Nevertheless, as we show in our experimental paper \cite{jafari2021experimental}, this incremental strategy leads to high computation costs. In \cite{liu2020ei}, authors extend I-LSH and introduce EI-LSH that features an aggressive early termination condition in order to stop the algorithm when good enough candidates are found and save the processing time. Considering that $R_{min}$ is the radius of the closest neighbor, $R$ is the current search radius in the original space, and $r$ is the current search radius in the projected space, EI-LSH changes $R_{min} \leq cR$ which is the stopping condition of I-LSH to $R_{min} \leq \lambda r$. Here, $\lambda$ is a pre-computed parameter that relies on the number of projections ($m$), the approximation ratio ($c$), and the success probability ($\delta$).

\noindent PM-LSH \cite{zheng2020pm} mentions that previous methods cannot estimate distances accurately because of using coarse-grained index structures, and as a result, they have to use larger search radiuses. Therefore, authors use PM-Trees to index the data and improve the query processing time. Moreover, PM-LSH uses a tunable confidence interval to use better distance estimations and offer a higher accuracy of the results. Authors in \cite{lu2020r2lsh} propose a novel two-dimensional method called R2LSH. Instead of using one-dimensional projections, R2LSH uses two-dimensional projections and maps dataset points into those projections. Furthermore, in the indexing phase, it builds B+-Trees for each of the two-dimensional projections. Later, in the query processing phase, R2LSH uses a query-centric ball to search the neighboring areas of the query and saves I/O costs.

\subsection{Angular-based LSH Techniques} % SUBSECTION

Super-bit LSH (SBLSH) \cite{ji2012super} focuses on the angular distance and mentions that previous methods suffer from a large variance in their angular similarity estimations. Therefore, SBLSH divides LSH random projections into multiple sub-projections, and then, it orthogonalizes multiple random projections for each sub-projection. The result of this strategy is several projections called super bits, and it is experimentally shown that using these super bits will result in a smaller estimation variance when the angle to estimate is within $(0, \pi/2]$. In \cite{ji2014batch}, authors propose batch-orthogonal locality-sensitive hashing (BOLSH) that uses batches of orthogonal projections instead of independent random projections for the angular similarity measure. Since these batches of projections partition the data space into regular regions, they improve accuracy. Further, authors show the benefit of BOLSH both experimentally and theoretically.

\subsection{Jaccard-based LSH Techniques} % SUBSECTION

LSH Forest \cite{bawa2005lsh} creates a prefix tree on each hash table and stores the compound hash keys in the prefix trees. In the query processing phase, a top-down and a bottom-up search is performed to find the points with the largest prefix match with the hash code of the query. This hierarchical search strategy makes it possible to stop in the middle of traversing the trees when enough results are found.

\subsection{Other LSH Techniques} % SUBSECTION
The following papers work with a combination of the hash function families or other application-specific families.

\subsubsection{Combination of hash function families:}\hspace*{\fill} \\
BayesLSH is proposed in \cite{satuluri2011bayesian}. The motivation of BayesLSH is that the false positives removal process in E2LSH \cite{datar2004locality} is expensive since Euclidean distance needs to be computed. Authors in this paper use the Bayesian statistics to find the probability distribution of similarities between the query and candidates by knowing the distribution of collisions in projections. This way, the Euclidean distance is estimated; thus, false positives removal process will be faster. BayesLSH can be applied to Euclidean, Angular, and Jaccard metrics, and authors show the benefit of their method both theoretically and experimentally. Authors in \cite{park2015fast} observe that LSH techniques such as E2LSH \cite{datar2004locality} and C2LSH \cite{gan2012locality} show poor performance on some data-sets while working well for the others. They mention that these techniques are significantly affected by the characteristics of the data-sets. In their paper, the hashed values of the points are called signatures and projection buckets are called signature regions. In some cases, the sizes of signature regions are different and points in a large signature region are not good candidates since they can be far from the query. Therefore, the proposed method (S2LSH) tries to distinguish between the signature regions and only use the important ones defined by two criteria: 1) smaller regions, and 2) regions that query is near center. Moreover, S2LSH can work with Angular and Euclidean distances.

\noindent In \cite{yu2016generic}, authors mention that LSH algorithms are slow (linear time complexity) when used in query workloads consisting of a large number of queries (e.g. using LSH to process similarity joins on two large data-sets). The main idea in this paper is to choose a small representative set from the query data-set to decrease the number of LSH lookups; thus, improve processing time. Several methods are proposed to solve the minimum set coverage problem (i.e. selecting representative query set) and the parameters of these methods are determined based on an error analysis on the final results. For the query processing, authors use an optimized version of QALSH \cite{huang2015query} which uses compound hash keys and R-trees. Moreover, authors show that their method can be applied to Euclidean-based, Jaccard-based, and Hamming-based versions of LSH.

\subsubsection{Kernelized hash function families:}\hspace*{\fill} \\
\cite{kulis2011kernelized} mentions that most of the LSH methods require the distribution and embedding of the input data to be explicitly known. In many scenarios, kernel functions are employed and the embedding of the data is not explicitly known. Therefore, authors propose a method to apply LSH functions to any kernel functions. The proposed method constructs LSH random projections using only a given kernel function and a sparse set of samples from the dataset. In \cite{chakrabarti2015bayesian}, authors improve on BayesLSH \cite{satuluri2011bayesian} by adding support for arbitrary kernel similarity measures. Authors use hyperplane rounding and sketch generation algorithms to adapt BayesLSH to kernel spaces and call their method K-BayesLSH.

\subsubsection{Application-specific hash function families:}\hspace*{\fill} \\
In \cite{lee2013locality}, LSH is modified to work on categorical data. The idea is to first find a similarity matrix between all categorical values, and then, use an agglomerative hierarchical clustering to cluster the categorical values. Then, each categorical value can be mapped to a cluster ID and the cluster IDs are projected to different buckets using LSH. In \cite{dong2019learning}, machine learning models are used to learn from dataset characteristics. In the offline processing phase, the NN search problem is mapped to a graph whose vertices are dataset points and each vertex is connected to its nearest neighbors. Then, a balanced graph partitioning algorithm is used to partition the graph into several bins such that the edges crossing between different bins would be as small as possible. Finally, a machine learning model is trained that given any point, it will predict the possible graph bins that the point can belong to. In the query processing phase, first, the graph bins of the query are predicted using the model, and then, the nearest neighbors are found from those bins only. The proposed method works with Edit and Optimal Transport distance metrics.

\section{Distributed Frameworks}
\label{sec:distributed}
In this section, we review the distributed frameworks that are proposed for LSH. There are several independent tasks happening in many LSH techniques (such as the multi-probing process in \cite{lv2007multi}) and researchers have used this motivation to build distributed frameworks for LSH techniques in order to further improve the performance.

\noindent \cite{Patil2007DistributedML} improved the time and space complexities using a distributed implementation of Multi-Probe LSH. The authors has developed their system using the master-slave architecture where the slave nodes are responsible to conduct the computational tasks, such as creating and maintaining the hash table, query searching, query ranking, and communication with the master node. On the other hand, the master node is responsible for splitting and distributing the large dataset, sending the query into the slaves nodes, and aggregating the ranked results. This work uses the master-slave architecture to search the LSH buckets and improve performance and scalability while maintaining the accuracy.

\noindent In \cite{haghani2009distributed}, a different distributed LSH architecture is proposed to improve the efficiency of KNN searches and range queries in high dimensional datasets. In order to map the LSH buckets to the peers of the distributed system, a two-level mapping strategy is used. This two-level strategy ensures that: 1) LSH buckets that are holding similar data are mapped to the same peers, and 2) load balancing between the peers is fair. The proposed method is followed by experiments to show that not only it meets the two requirements but also minimizes the network I/Os. Layered LSH, proposed in \cite{bahmani2012efficient}, uses Apache Hadoop for its disk-based version and Active DHT for its in-memory version. Authors provide theoretical guarantees for the network operations in the single hash tables setting. However, for the multiple hash tables setting, no theoretical guarantees are provided. Layered LSH works in the Euclidean space and uses Entropy LSH \cite{panigrahy2005entropy} as its base LSH method.

\noindent Parallel LSH (PLSH) \cite{sundaram2013streaming} introduces an in-memory, multi-core, distributed variant of LSH that can be used to perform KNN searches on large and streaming data. PLSH, uses a caching strategy to improve the online index construction time of the streaming data. Moreover, it uses insert-optimized delta tables to hold the indexes of new incoming data while merging them with the main index structures periodically. Furthermore, a bitmap-based strategy is used to eliminate duplicate data fetched from different hash tables. Additionally, PLSH is designed to work only on the angular distance.

\noindent An LSH-based distributed framework is proposed in \cite{karapiperis2013distributed} that improves scalibility in the privacy preserving record linkage domain. The framework utilizes the Map-Reduce paradigm and uses LSH to find the Minhash signatures of the records in the Map phase. The signatures are then encoded into Bloom filters and the Bloom filters are distributed over the Reduce tasks.
CLSH \cite{xu2014clsh} uses the K-Means clustering algorithm to split the original dataset into multiple clusters. Later, it distributes these clusters over different compute nodes where each one creates their local indexes. In the query processing phase, the given query is compared to the cluster centers and nearest compute nodes based on the Euclidean distance metric are chosen to run LSH on them. Finally, the intermediate results are combined and the nearest points are chosen as the final results.

\noindent In \cite{li2017ses}, the authors introduce a naive distributed version of LSH on Apache Spark called SLSH. In the indexing phase of SLSH, each worker node loads a subset of the dataset and calculates the hash values of the points in that subset. Later, in the query phase, all worker nodes load the hash functions and hash tables and the query is sent to all of them for query processing. The main downside of SLSH is that it requires several data shuffles across the worker nodes that result in heavy network costs. To overcome this issue, authors present a more efficient version called SES-LSH. SES-LSH uses a hashing scheme (called BKDR) to partition data such that points belonging to the same hash tables are partitioned to the same worker nodes. Therefore, since the location of data points is known in advance, computations can be performed locally and it is not required to send the query to all worker nodes.
\cite{li2017losha} proposes a generic distributed platform, called LoSHa, that can be used to easily implement distributed versions of different LSH methods with different distance metrics. LoSHa can lower programming costs and also achieve high efficiency and performance. Internally, LoSHa uses the Map-Reduce paradigm and offers several user-friendly APIs to ease the process of converting an LSH algorithm into a distributed version. Furthermore, by using different optimization techniques such as bucket compression and data de-duplication, LoSHa improves the efficiency of the implemented method.

\noindent C2Net \cite{li2018c2net} focuses on the collision counting operation of LSH methods such as C2LSH since its authors believe collision counting is the most time consuming operation compared to hash value calculation and false positive removal operations. C2Net utilizes minimum spanning trees and spectral clustering to partition LSH buckets and distribute them over mapper tasks in a Map-Reduce framework. Moreover, C2Net supports virtual rehashing by running two rounds of Map-Reduce and merging different bucket blocks.
\cite{shen2018towards} and its extended version \cite{wu2020lsh} focus on the problem of load balancing in the indexing phase of distributed LSH-based methods. They propose two theoretical models that can predict the data distribution of a single hash table. Later, using the two theoretical models and CDF-based and virtual node methods, a dynamic load balancing strategy is proposed. Finally, the proposed method is followed by experimental evaluations that use the Gini coefficient metric in order to prove the scalability.
A distributed approach, named RDH, is proposed in \cite{durmaz2019fast} to improve the speed of performing similarity search for images. RDH randomly splits and distributes the dataset over different compute nodes and each node runs LSH indexing locally using their local subset of the data. Furthermore, authors show that the accuracy does not change significantly if the same hash functions are used in all compute nodes.

\section{Applications}
\label{sec:applications}
In this section, we present different works in diverse application domains that utilize Locality Sensitive Hashing. Note that, we counted more than 1000 application papers that utilize LSH in their application workflow. Here, we have carefully categorized the most recent and popular papers (in terms of their citations). For each paper, we present the particular problem that they are trying to solve, and how LSH is utilized in their methodology. Additionally, we also note the specific LSH algorithm and hash family that is used by the paper (if the paper mentions it). The goal of this section is to present the reader the knowledge of how LSH is used in these diverse domains. 

\subsection{Audio Processing}

\cite{ryynanen2008query} presents a query by humming method for music retrieval using LSH. The pitch vectors are extracted for a music database and an index structure is constructed. For retrieval, the query transcription technique is employed to produce notes for the song, and then, pitch vectors are extracted similar to the pitch vectors extracted in the index construction phase. Later, the E2LSH \cite{datar2004locality} method is used to return the nearest neighbors to finalize the list of similar songs. Moreover, the Euclidean distance metric is used to find the distances between the pitch vectors.

\noindent In \cite{yu2010combining}, Order Statistics LSH (OS-LSH) is proposed to improve the scalability of content-based retrieval of audio tracks in music databases. For an audio query, Chroma sequences are calculated and then a Multi-Probe histogram (MPH) is generated for the sequences. Later, OS-LSH maps MPH into hash keys. Finally, in the post-filtering step, the histograms associated with the keys in the hash table are compared against the query and similar items are returned. The representation using histograms make the proposed method more storage-efficient and scalable.

\noindent \cite{han2011content} solves the computational complexity of finding similar musics from their content-based extracted features using LSH. After extracting the Mel-frequency Cepstral Coefficients (MFCC) and Time Histogram (TH) features from the music, the K-Means clustering algorithm is performed to have a Bag of Words (BoW) representation. Later, LSH is used to find similar clusters instead of a linear search approach which would otherwise take O($N^2$).

\noindent \cite{yu2013scalable} speeds up the retrieval of similar songs based on melody similarity in a large database. Initially, a Support Vector Machine (SVM) model is trained by taking the generated Chord Progression (CP) into consideration for a set of audio tracks. The Chord Progression Histograms (CPH) are computed for the audio tracks and organized in to one single hash table with a tree-structure while considering CP as the hash key. In the query processing phase, the CPH of CPs are computed and similar songs are retrieved from the corresponding hash buckets.

\noindent \cite{padmasundari2017raga} proposes an identification method that uses LSH for Raga which is a quintessential component of classical Indian music. The new music recordings are stored in the database. Then, the pitch vectors are extracted and stored along with their labels. Later, LSH is used to index the pitch vectors. Similarly, the pitch vectors are extracted for the query set and then compared with the indexed pitch vectors using the Euclidean distance metric in order to find the Top-k Raga to the query.
\subsection{Image/Video Processing}

In \cite{kulis2009kernelized}, the authors extend the scope of LSH to arbitrary kernel functions while preserving the algorithm's sub-linear time and propose KLSH. In KLSH, the random projections are computed for LSH which are required only in the kernel space and for a limited number of database objects to find the set of similar images to the query. Moreover, KLSH utilizes Gaussian RBF kernel to retrieve the images.
		
\noindent \cite{zhang2016video} introduces a new approach for video anomaly detection using LSH filters. The training data points are hashed into a set of buckets using LSH. With the help of bloom filters, a test data point will be detected as abnormal if it falls into a different bucket and will be normal if it falls into the same training bucket. The hamming distance measure is used to find the distances between the training buckets and the test bucket. Furthermore, The Practical Swarm Optimization (PSO) method is also used to search for optimal hash functions which further improves the detection quality.
		
\noindent In \cite{xia2016privacy}, the authors propose a method to support content-based image retrieval over encrypted images in cloud applications. Initially, feature vectors are extracted for the corresponding input images and the k-NN method is employed for the encryption of feature vectors. Then, pre-filtered tables are created using LSH. E2LSH \cite{datar2004locality} is used to construct the tables. In order to further enhance the security, watermark-based protocols are used to prevent the illegal distribution of images by the legal users. Furthermore, Euclidean distance between the feature vectors is used to find the similarity of images.

\subsection{Security/Privacy Related}

\cite{opricsa2014locality} accelerates the clustering process of large malware datasets by using the LSH technique. The malware samples are represented as a set of features and the distance between the two samples is computed using the Jaccard distance. Then, minHash functions are computed efficiently to find the set of similar malware samples by using the banding approach. Later, the malware samples are clustered by using the Single Linkage (SLINK) algorithm.

\noindent \cite{ozawa2015online} introduces a new online system to detect malicious spam emails by using a Resource Allocating Network and LSH (RAN-LSH). LSH is used to select the training data that has to be learned by the RAN-LSH classifier to detect the spam emails. For the test data, the hash table is looked upon to find same or similar spam emails.

\noindent In \cite{zhang2016scalable}, the authors improve the scalability of local recoding (a technique used to anonymize data and preserve privacy) in big data applications. A semantic distance metric is proposed in order to measure the similarity between data points. Later, Minhash LSH and Map-Reduce are used to split the data into several partitions that contain similar records. The anonymization of these partitions is done via a recursive agglomerative k-member clustering method.

\noindent In \cite{azimpourkivi2017secure}, the authors introduce a novel authentication system, called \textit{ai.lock}, for mobile devices which uses an imaging sensor for authentication. To extract invariant features for image-based authentication, LSH is used along with deep neural networks and Principal Component Analysis (PCA). The architecture processes the input image through neural networks and LSH is employed to map them to a binary image print. It also uses a classifier to identify the ideal error tolerance threshold to lock and match image prints. Moreover, this work uses the hamming distance metric.

\subsection{Blockchain}

\cite{zhuvikin2018blockchain} presents a blockchain scheme for image copyrights and provides the copyrights over the network under distribution constraints. First, the image feature vector is calculated to represent the image content for the input images; then, the feature vector is added to the blockchain. When a user wants to use the photo/image which is present in the blockchain network, E2LSH \cite{datar2004locality} is employed to find out the copyright owner by using the information such as authority item, image index item, usage item, and ownership change item present in the network as data blocks.
		
\subsection{Data Mining Approaches}

In \cite{ravichandran2005randomized}, the authors reduce the running time of similarity list creation for nouns gathered from a web corpus. Nouns clustering is a well-known task in Natural Language Processing and creating the similarity matrix is an expensive operation of noun clustering. Therefore, authors use LSH and cosine similarity metric to create an approximate similarity matrix and speedup the operation.

\noindent \cite{koga2007fast} uses LSH in the single-link hierarchical clustering technique to approximate the  distances and reduce the time complexity of finding nearest clusters. Authors experimentally and theoretically show that using LSH in their proposed method reduces the time complexity significantly. It is also worth mentioning that Euclidean distance is used as the distance metric.

\noindent Often, researchers use sampling techniques while removing outliers from the initial sample to perform association rule mining on large datasets in a reasonable amount of time. \cite{chen2011locality} mentions that the initial sample may contain multiple clusters and performing data clustering on the initial sample can result in an increase in the accuracy. Therefore, authors use LSH and Euclidean distance metric to first cluster the initial data sample, and then, remove outliers from the buckets.

\noindent LSHiForest \cite{zhang2017lshiforest} uses LSH forest to propose a framework for ensemble anomaly analysis. LSHiForest is built upon iForest, which is an isolation based anomaly detection forest. Moreover, the proposed framework has the ability to use any distance metric and any type of LSH family such as Euclidean-based, angle-based, and kernelized LSH. Finally, authors experimentally show that LSHiForest beats other methods in terms of time efficiency, anomaly detection quality, and robustness.

\noindent Direct Robust Matrix Factorization (DRMF) is a technique used in the anomaly detection domain. \cite{xie2017fast} argues that although DRMF is robust and accurate; however, it involves expensive computations. To speedup traffic anomaly detection, \cite{xie2017fast} proposes a multi-layer LSH table that maps origin and destination pairs to different layers with different similarity levels (based on the Euclidean metric). Later, an adaptive strategy is proposed to search the generated layers.

\subsection{Text/Document Processing}

\cite{jiang2011semi} presents a new method called Semi-Supervised SimHash $(S^3H)$ to search for similar documents in high-dimensional spaces. Since it is semi-supervised, it learns the optimal feature weights and the weights are used to find the query results since similar objects have similar fingerprints. Initially, the data set is mapped into an L-dimensional Hamming space using LSH, and then, the fingerprints are generated which are used to find the similar documents.
		
\noindent In \cite{garcia2019locating}, the paper proposes a method to identify misspelled names and near-duplicates using LSH. First, the data is transformed and similar names (candidates) are produced from the transformed data using LSH with Jaccard distance. The candidate pairs are then filtered using Full Damerau-Levenshtein distance. Similar names are aggregated into a set of names by utilizing a graph.
		
\noindent \cite{li2014large} proposes a method for large-scale document reduction based on domain ontology and LSH. In the first step, the features are extracted using a Semantic Vector Space Model (SVSM). Then, the index of SVSM is obtained using E2LSH \cite{datar2004locality}, and then, each document is mapped into the hash tables. Later, the candidate set of similar documents is obtained by getting the union of the buckets that have similar documents. Lastly, Euclidean distance between the query and the documents in the candidate set are computed to retrieve the true similar documents.
		
\noindent \cite{van2010online} generates online LSH signatures in order to process large text collections. The authors improve upon offline generation of LSH signatures and propose an algorithm that is suitable for streaming applications. It is also space-efficient since the method does not need an explicit representation of the feature vectors or random matrices. For every feature, it uses a fixed pool of random values rather than creating a unique value for each feature. Furthermore, the proposed method uses cosine similarity of the feature vectors.
        
\subsection{Biological Sciences}

LSH-ALL-PAIRS \cite{buhler2001efficient} efficiently compares genomic DNA sequences with the goal of finding conserved genome features across different species. LSH-ALL-PAIRS converts the sequences to shingles. Then, LSH is applied to all shingles and the shingles with the same hash value are grouped together into a class. Finally, a pair-wise comparison is performed only on a specific class to find similar shingles. \cite{berlin2015assembling} introduces MinHash Alignment Process (MHAP) to detect overlaps between noisy and long reads of microbial genomes. MHAP uses MinHash to create small fingerprints of sequencing reads with the goal of dimensionality reduction. To do this, MHAP decomposes DNA sequences into multiple shingles, and then, the shingles are converted into integer fingerprints using multiple random hash functions. Finally, the Hamming distance of fingerprints is used to approximate the Jaccard distance of two shingles that helps determine the overlaps between them.

\noindent MASH \cite{ondov2016mash} facilitates the use of MinHash in data-intensive problems in genomics by proposing a general-purpose toolkit to construct, manipulate, and compare MinHash fingerprints from genomic data. Moreover, MASH derives a significance test and proposes a new distance metric, called Mash distance, that estimates the mutation rate of two sequences. Mash distance can easily be computed using only the MinHash fingerprints.

\noindent Molecular fingerprints are often used to describe, compare, and benchmark organic molecules. MinHash Fingerprint up to six bonds (MHFP6) \cite{probst2018probabilistic} is a fingerprint that adopts LSH to improve the performance of nearest neighbor searches in benchmarking studies. In order to do this, MHFP6 applies MinHash to molecular substrings and generates multiple fingerprints. The fingerprints are then indexed by LSH Forest \cite{bawa2005lsh} to efficiently retrieve nearest neighbors.

\subsection{Geological Sciences}

\cite{yu2013edge} improves the scalability of geo-fencing applications by processing hundreds of polygons and points in real-time. Initially, an R-tree is used to quickly detect whether a point is present in a minimum bounding rectangle. Then, an edge-based LSH technique is used in large-scale pairing between points and polygons for INSIDE and WITHIN detection of points followed by a probing method to find out all the geo-edges close to a target point.
		
\noindent In \cite{rantanen2015speeding}, the authors speed up the construction of roadmaps without leveraging the quality by using LSH. Centroid-based hashing is employed to search for nearest neighbors during the construction phase. The centroids are initialized first and each centroid corresponds to a region of a Voronoi cell. An arbitrary point is associated with one of the centroids by calculating the Euclidean distance of the point to all of the centroids and then selecting the nearest distance. To retrieve a set of nearest points, it is sufficient to retrieve it from the points in the Voronoi cells.
 
\noindent In \cite{moura2017lshsim}, the authors speed up the process of searching for patterns similar to a target one in 2D and 3D image training libraries. To efficiently search for a pattern in training images, LSH is used first to filter the patterns that are similar to a given data event. Then, an exhaustive search using a Run-Length Encoding (RLE) compression technique is used to calculate the similarity among the filtered patterns. The Euclidean distance measure is used for continuous images and the Hamming distance measure for categorical images.

\noindent \cite{astefanoaei2018multi} reduces the computation costs of nearest neighbor search, distance estimation, clustering, and classification of GPS trajectory data. LSH is applied to two distance measures named Hausdorff and Fréchet distances which are used in the neighbor search. Furthermore, a data structure called the Multi-Resolution Trajectory Sketch (MRTS) is built to compactly represent the dataset. This also helps for fast insertion of trajectories in the database.
        
\subsection{Graph Processing}

\cite{dutta2013symbol} utilizes LSH and proposes a symbol spotting approach in graphical documents. The proposed method uses the critical points of graphical documents as the nodes and the lines joining those critical points as the edges of a graph. Later, the graph is decomposed into multiple graph paths, and finally, the shape descriptors of the graph paths are mapped to hash tables using LSH. 

\noindent \cite{zhang2015fast} focuses on similarity search in undirected vertex-labeled simple graphs that have no self loops and multiple edges. Given a query graph with a semantic class label, the goal is to find the nearest graphs that have the same class as the query graph. The authors propose a vectorial representation method that is used to convert the graphs to high-dimensional vectors. Finally, Multi-Probe LSH \cite{lv2007multi} and Euclidean distance are used to query the generated high-dimensional vectors.

\noindent Basic Weighted Graph Summarization (BWGS) is a method used to compress graphs that requires finding similar nodes within a graph. \cite{khan2017faster} uses Min-Hash to approximately find the required similar nodes, and as a result, speedup the process of graph compression. Moreover, for graphs that contain edge weights, a Weight Oriented LSH (WOLSH) strategy is proposed to increase the chance of generating similar Min-Hash values from a subset of neighbors that have higher edge weights.

\subsection{Machine Learning}

Maximum Inner Product Search (MIPS) is the problem of finding a dataset point that has a maximum inner product to a given query point. Mathematically, this problem can be converted to a near neighbor search problem when the norm of every dataset point is constant. However, this is not the case in many applications and the MIPS problem cannot easily be solved by using near neighbor search techniques. In \cite{shrivastava2014asymmetric}, authors propose an asymmetric LSH (ALSH) method that uses different hash functions for buckets creation and buckets probing. ALSH can be easily applied to MIPS to improve the performance, and it is based on E2LSH \cite{datar2004locality} that uses the Euclidean distance metric.

\noindent In \cite{chen2016structural}, the authors solve the scalability issue of applying large trained models on huge non-annotated media collections. A trained linear Support Vector Machine (SVM) classifier requires a weight vector, features vector, and a bias in the formula $h(x)=sgn(w.x+b)$ to classify the features vector. As a result of using LSH, the approximated $h(x)$ value can be found by a range query in the Hamming cube anchored around the hashed value of $w$. Hence, they build an approximated classifier using LSH by hashing the weight vector and the features while knowing that the dot product of $w$ and $x$ can be estimated by the Hamming distance of their hashed values.
 		
\noindent Automatic Speech Recognition (ASR) is a method used to convert speech into text. One strategy to improve the performance of ASR is to apply it on features derived from a manifold learning based approach. However, this process requires computing pair-wise distances between the feature vectors to construct nearest neighborhood graphs that are required in manifold learning techniques. LPDA-LSH \cite{tomar2013efficient} is proposed to solve this issue using a modified version of LSH. In this modified version, pairwise distances between all hashed values are calculated for each hash bucket with the goal of creating candidate sets for all the points in that bucket. Later, the candidate sets are concatenated based on class labels and a within-class and an inter-class neighborhood graphs are created.

\noindent \cite{yan2018norm} mentions that the state-of-the-art hashing based technique for MIPS uses a normalization strategy with the maximum 2-norm in the dataset and this strategy suffers from performance issues when used in real datasets that have long tails in their distributions. Therefore, authors introduce NORM-RANGING LSH that divides the dataset based on the percentiles of the 2-norm distribution. Later, the state-of-the-art technique is applied individually to each dataset partition. Moreover, a new similarity metric is proposed to define how to probe from different partitions of the dataset.

\noindent H2ALSH \cite{huang2018accurate} presents a novel transformation method to reduce the error that is caused by asymmetric strategies applies on MIPS. The novel transformation method is called Query Normalized First (QNF) and converts the MIPS problem to a nearest neighbor search problem. H2ALSH first divides the dataset into multiple subsets and uses QNF to transform the subsets. Later, it uses QALSH \cite{huang2015query} to build the indexes for the subsets. Finally, the union of the results from the subsets are reported as the MIPS result.

\subsection{Healthcare}

Stratified LSH (SLSH) \cite{kim2016stratified} presents a detection system for high-dimensional physiological data using LSH. SLSH which is a type of multi-level LSH predicts the critical events of a patient. First, the training data is stratified with the help of LSH by using $l_{1}$ distance. Then, Cosine distance LSH (COSLSH) is applied at the inner level on each bucket. To retrieve the approximate nearest neighbors of the query, the same two levels of LSH are applied and a linear search is performed within the candidate set. Finally, the prediction is made by the majority vote technique.

\noindent \cite{xu2018bone} demonstrates an LSH-based approach to retrieve bone scan images by using the SIFT-based Fly Locality Sensitive Hashing (FLSH) technique. First, the Difference of Gaussian (DOG) is calculated for the input images to detect potential minimum points as key points. Then, the value of normalized Laplacian function at each key point is evaluated and is assigned one or more orientations if it meets a certain threshold. Finally, a SIFT 128-dimension feature vector is generated for each image and hash codes are produced for the feature vector.

\noindent \cite{alotaiby2019locality} improves the prediction accuracy of critical events for a patient in a medical database. In the indexing phase (also called training step) of the proposed method, LSH is used to hash and index the features that are extracted from ECG signals of a subject. In the query processing phase (also called testing step), the extracted features of a new ECG signal are also hashed into the LSH buckets, and the class of this new signal is determined based on the class of the majority of its near points.

\subsection{Plagiarism/Near-Duplicate Detection}

\cite{chum2007scalable} presents two novel schemes for near-duplicate image and video shot detection. The first scheme uses the hierarchical tiled color histogram for image representation, and it uses Euclidean distance to compute the similarity. Then, image retrieval is achieved using LSH. The second approach uses a sparse set of visual words for image representation. Then, the set overlap measure is used to compute the similarity and the Min-Hash algorithm is employed for efficient retrieval of images.

\noindent SimPair LSH \cite{fisichella2010efficient} solves the problem of near-duplicate detection for high dimensional data points incrementally and efficiently. Initially, a certain number of pair-wise similar distance sets that meet a threshold for existing data points are stored in the memory. Later, in the query processing phase, if one of the points from a similar pair appear in the same bucket as the query, it is very probable for the other point to also appear in the bucket. Therefore, SimPair LSH can avoid computing distances for the points that are similar to each other at the first place; thus, it can save the overall processing time. Additionally, SimPair LSH is an in-memory index structure that uses Euclidean distance metric.

\noindent \cite{sood2011probabilistic} presents a technique to detect similar documents using SimHash. Probabilistic SimHash Matching (PSM) is proposed that incorporates a proposed algorithm called Volatility Ordered Set Heap (VOSH). VOSH randomly flips the bits without repetitions. PSM finds similar documents using online and batch operation modes. The online mode of operation is adopted for the documents which fit in to the main memory and the batch operation mode is preferred for the documents that are kept on the disk.

\noindent \cite{al2016copy} proposes a method for Copy-Move Forgery detection of images using K-Means clustering and LSH. First, the image is divided into non-overlapping blocks and features are extracted from these blocks. Later, a vector array is formed from these features and the PCA method is used for dimensionality reduction of vectors into two dimensions. In the next step, the reduced vectors are clustered into multiple clusters using the K-Means clustering method. Finally, LSH is deployed to find the matching pair of blocks based on the Euclidean distance, which in turn, results in a list of candidate pairs for the forgery.
	    
\noindent In \cite{hu2018deep}, the authors introduce an efficient method for near-duplicate detection using a neural network and a load-balanced LSH approach. The neural network extracts the features for the detection process. Then, LSH is used to build an index for the extracted features. A load-balanced LSH method is proposed to map images into buckets in a balanced manner and find the relevant number of neighboring buckets to detect duplicates. The Load-balanced LSH uses the Euclidean distance metric.

\subsection{Networks}

\cite{hou2008semantic} integrates a semantic searching method based on LSH in Mobile P2P networks. Initially, for a document vector, the semantic indexing method is used to hash the document vector into a key by using entropy-based LSH \cite{panigrahy2005entropy}, and it stores the key in the associated mobile node (mobiles, laptops, etc.) in the network. Moreover, there are some super nodes, called stationary nodes, present in the network which act as access points. When any of the mobile nodes want to query the document, query is sent to access points, and then, super nodes are responsible to choose various points randomly from the neighborhood of the query.
		
\noindent NearBucket-LSH \cite{kraus2016nearbucket} solves the similarity search problem in P2P online social networks. Initially, LSH is used to map the users into a collection of buckets. Therefore, when a query is received, LSH limits the searching process to the buckets to which query is mapped. NearBucket-LSH is based on Multi-probe LSH \cite{lv2007multi} and also uses Content Addressable Network (CAN). CAN allows to map the users to the buckets and store the buckets in distributed nodes across the network. Furthermore, CAN is used to update and locate the required buckets in the query processing phase.

\noindent In \cite{ali2016neural}, the authors use a neural network model to detect Distributed Denial of Service (DDoS) cyber attacks by using the Resource Allocating Network with LSH (RAN-LSH) classifier. First, in the pre-processing step, the transformation of darknet packets to the feature vector is carried out. In order to train the neural network, LSH is used to select only a certain amount of data to be trained for the model which in turn accelerates the learning time of the model.

\noindent \cite{tian2020lsh} presents a network congestion detection method for the Signal Safety Data Network (SSDN). Initially, the data flow of SSDN is decoded and fed into LSH as input. LSH is used to determine whether there is network congestion or not. If the hash buckets overflow, then there is a network congestion and vice versa. If there is a congestion in the network, the data is pre-processed by extracting the features and then normalizing it. Later, the XGBoost algorithm is used to determine the congestion type of the pre-processed data.

\subsection{Software Testing}

Tree Locality-Sensitive Hash (TLSH) \cite{cady2017tree} uses Min-Hash to map trees that are similar to each other to the same hash value using the tree edit distance metric. TLSH is applied to the path constraints of symbolic execution states of software to identify the similar states and help to find the bugs more efficiently.
  
\subsection{Ontology Matching}

\cite{cochez2014locality} uses Min-Hash to match equivalent terms in different big ontologies. Class labels are used to create a string-based alignment. Two strategies are adopted to make the alignment usable in Min-Hash. In the first strategy, class labels are shingled and a hierarchical method is used to merge the shingles. In the second strategy, a class is represented using the different tokens from the class label. The representation is then fed into the Min-Hash algorithm. Finally, two compare two ontologies, one of them is used as the dataset and the other one is used as the query.

\noindent \cite{cochez2017large} uses LSH Forest \cite{bawa2005lsh} to find the right context of the environment for placing new knowledge tokens. Moreover, the Random Hyperplane Hashing (RHH) method is used along with LSH Forest to utilize the angular distance metric. 

\subsection{Social Media and Community Detection}

\cite{da2013similarity} uses Vector Space Models (VSM) to represent user profiles in social media and Random Hyperplace Hashin (RHH) to create an index for the user profiles. RHH is an LSH family that uses cosine distance. Moreover, authors experimentally show the benefits of using RHH to develop similarity search systems for social media profiles.

\noindent Tag Assignments Stream Clustering (TASC) \cite{wu2014incremental} is a method that uses LSH for community detection in social tagging systems. To detect the communities, TASC generates user profile vectors based on users' interests. Later, LSH and angular distance is used to find the similarities between the users and create user clusters. In real-time systems, the clusters are updated over time and similarities between new users and the current users are recalculated.

\noindent \cite{abdulhayoglu2018use} proposes an LSH-based method to match papers in Web of Science and Scopus bibliographic databases. First, LSH is used to find similar papers from the two databases in a reasonable amount of time. Later, heuristic-based approaches are used to remove the false positives and obtain the exact matches. The utilized LSH method uses cosine distance metric and is implemented on Spark to further improve the speed by distributed processing.

\subsection{Time Series}

Authors in \cite{kim2016analysis} use two LSH methods to predict critical events from physiological time series. L1LSH \cite{gionis1999similarity} that is based on Hamming distance and E2LSH \cite{datar2004locality} that is based on Euclidean distance are used to find the top nearest neighbors to a given query. Later, the class label of the query is chosen using a majority vote of its nearest neighbors.

\noindent \cite{kim2017physiological} mentions that using a generic LSH method for physiological time series has two problems: 1) not being able to use multiple distance metrics at a time, and 2) expensive distance computations on the candidates set in order to remove the false positives. Therefore, in their thesis, the author proposes Stratified LSH (SLSH) to solve the first problem and Collision Frequency LSH (CFLSH) to solve the second problem. SLSH uses a multi-level hierarchical approach that supports using different distance metrics at each level. CFLSH uses a collision counting strategy with the intuition that the more similar a point is to a given query, the more times it will collide with the query across multiple hash tables.

\noindent In \cite{rong2018locality}, authors use LSH to identify potential earthquakes by finding similar seismic data time series. First, a fingerprint extraction strategy is used to convert the input time series into compact binary vectors. Later, Min-Hash is applied to the generated binary fingerprints in order to identify all similar fingerprints. Moreover, authors also study the effect of parallel processing (using multiple CPU threads) on the processing times.

\noindent \cite{dhamala2018multivariate} reduces the cost of detecting an Acute Hypotensive Episode (AHE) given the vital signals of ICU patients in the form of multivariate time series using LSH. First, Bidirectional Sequence-to-Sequence (BSS), Hierarchical Sequence-to-Sequence (HSS) autoencoders, and a combination of them is used to encode the input time series into context vectors. Later, Stratified LSH (SLSH) \cite{kim2016stratified} is used to find the similar context vectors. 

\noindent \cite{yu2019fast} mentions that finding similar time series is gaining importance these days due to advances in mobile devices and sensors. Therefore, authors use LSH to find candidate similar time series, and then use the hash values to estimate the original distances and prune the candidate sets. Moreover, authors find the appropriate LSH parameter by performing an error analysis. The proposed method uses QALSH \cite{huang2015query} and both Dynamic Time Warping (DTW) and Euclidean distance metrics.

\subsection{Robotics}

\cite{tanaka2008scalable} improves the scalability of the Monte Carlo Localization (MCL) algorithm that is used for global localization and position tracking in robotic systems. E2LSH \cite{datar2004locality} is employed to select the features that are required to build the scalable MCL framework. For building the features database incrementally, the mapper robot gets the new feature and hashes the features using E2LSH and adds the location to the corresponding buckets.
		
\noindent LSH-RANSAC \cite{saeki2009lsh} solves the problem of feature-based robot localization in large-size maps. For appearance based localization, incremental maps based on iLSH \cite{tanaka2008scalable} is used to build an incremental database. For a new feature, the real-world location is computed, and then, the feature is hashed using E2LSH \cite{datar2004locality}. Then, the real-world location of the feature will be associated with the hashed values. For position-based localization, iRANSAC \cite{tanaka2006incremental} is employed which is a map-matching scheme. Both of the iLSH and iRANSAC techniques are used to solve real-time localization problems.
		
\noindent In \cite{heise2015fast}, the authors improve the processing speed of high-resolution stereo images in robotics. To establish dense image correspondences, the approximation of one image with another image has to be computed. LSH is used to solve the approximation problem by using the dense binary strings of the image pixels.
	
\section{Conclusions}
\label{sec:conclusions}
In this survey paper, we reviewed the recent advances in Locality Sensitive Hashing (LSH) techniques and categorized them based on the hash function families that they utilize. Additionally, we reviewed the distributed frameworks proposed for LSH techniques and explained their architecture. Finally, we categorized different application papers and presented how Locality Sensitive Hashing is utilized in each of them.

\bibliographystyle{ACM-Reference-Format}
\bibliography{references}

\end{document}